\documentclass[epj,twocolumn]{webofc}
\usepackage{color}
\def\ee{\end{equation}}
\def\be{\begin{equation}}
\def\eea{\end{eqnarray}}
\def\bea{\begin{eqnarray}}
\def\bal{\begin{align}}
\def\eal{\end{align}}

\def\be{\begin{equation}}
\def\ee{\end{equation}}
\def\bea{\begin{eqnarray}}
\def\eea{\end{eqnarray}}

\def\bef{\begin{figure}}

\def\eef{\end{figure}}
\woctitle{ISVHECRI 2014}
\begin{document}
\title{Photoproduction models for total cross section and shower development}
\author{Fernando Cornet\inst{1}\fnsep\thanks{\email{cornet@ugr.es}} 
\and
 Carlos Garcia Canal \inst{2}\fnsep\thanks{\email{cgarciacanal@fisica.unlp.edu.ar}}
  \and
                Agnes Grau \inst{1}\fnsep\thanks{\email{igrau@ugr.es}} \and
Giulia Pancheri\inst{3}\fnsep\thanks{\email{pancheri@lnf.infn.it}}\and 
        Sergio Sciutto \inst{2}\fnsep\thanks{\email{sciutto@fisica.unlp.edu.ar}}
}

\institute{Departamento de F\'{\i}sica Te\'orica y del Cosmos and
Centro Anadaluz de F\'{\i}sica de Part\'{\i}culas, Universidad de
Granada, E-18071 Granada, Spain
\and 
           Departamento de F\'{\i}sica, Universidad Nacional de La
Plata, IFLP, CONICET, C.C.67, 1900 La Plata, Argentina
\and
INFN Frascati National Laboratories, Via E. Fermi 40, Frascati, Italy 00044
          }
          \abstract{%
  A model for the total photoproduction cross section
based on the ansatz that  resummation of  infrared gluons  limits  the rise 
induced by QCD minijets  in all the total cross-sections, is used to simulate extended  air showers initiated by cosmic rays  with the AIRES 
 simulation program.
The impact on common shower observables, especially those related
with muon production, is analysed and compared with the
corresponding results obtained with previous photoproduction models.
}
\maketitle

\section{Introduction}
Cosmic rays have often anticipated discoveries in particle physics and have allowed exploration of energies which were  not available by particle accelerators at the time. Such was the case of the rise of the total $pp$ cross-section in the early '70s \cite{Yodh}, an observation confirmed shortly after by ISR \cite{ISRxsect}. The rise of all hadronic cross-sections was soon suggested  to be due to an increasing number of parton-parton collisions \cite{clinehalzen}, which can be  described   by perturbative QCD  and are referred to as mini-jets. An early application of an eikonalized mini-jet model for cosmic ray interactions  \cite{durand1987} still reproduces within errors the recent AUGER measurement of the total $p-air$ cross-section \cite{AUGER2012}. Different models for the basic $pp$ interaction   and its applications to cosmic rays are discussed in the literature, and included in MonteCarlo simulations, many of them using mini-jets as the driving phenomenon of the rise. Mini-jet models have been extended  to photoproduction, with some additional parameterisation of the parton content of the photon, e.g. in  \cite{halzenfletcher,block1999} and \cite{ourEPJC}.

In this contribution  we shall study shower observables in cosmic ray in photo production  obtained by  two different models for the total $\gamma -p$ cross-section as input to the AIRES MonteCarlo \cite{AIRES},  the eikonalized mini jet model with soft gluon resummation of \cite{ourEPJC} and  the analytic amplitude model by Block and Halzen \cite{block2014}, which is presently the one used in AIRES. We shall compare predictions at very high energies, where QCD processes are expected to be dominant, and highlight the differences in different observables.
 \section{ Photoproduction mini-jet model for total cross section}
In order to study $\gamma-air$ at very high energies, we first  update the results of the eikonal mini-jet model for $\gamma-p$ interaction of \cite{ourEPJC} in light of the recent LHC results for the total $pp$ cross-section \cite{TOTEM7,TOTEM8,ATLAStot}.

We shall start with a brief summary of the content of the model for $pp$ and its extension to $\gamma-p$.

Mini-jet models for  the total cross-section use as input  a perturbative calculation based on the QCD jet cross-section, namely
\begin{align}
 \sigma^{AB}_{\rm jet} (s,p_{tmin})=
\int_{p_{tmin}}^{\sqrt{s}/2} d p_t \int_{4
p_t^2/s}^1 d x_1  \int_{4 p_t^2/(x_1 s)}^1 d x_2 \times \nonumber \\
\times \sum_{i,j,k,l}
f_{i|A}(x_1,p_t^2) f_{j|B}(x_2,p_t^2)
  \frac { d \hat{\sigma}_{ij}^{ kl}(\hat{s})} {d p_t}
  \label{minijets}
  \end{align}
with $A,B = p, \bar p,\gamma$.   The parton-parton differential cross-section  $d \hat{\sigma}_{ij}^{ kl}(\hat{s}) /d p_t$   defines the total contribution for partons with $p_t>p_{min}$,  where the     parameter
$p_{tmin}
\approx 1-2$ GeV separates   hard processes
for which one can use a perturbative QCD description, from the
soft ones which dominate at low c.m. energy of the scattering hadrons,  $\sqrt{s} \lesssim 10\div 20\ GeV$. The mini-jet cross-section gets its name  because is dominated by low-$p_t$  processes, which cannot be identified by jet finding algorithms, but  can still be perturbatively calculated using parton-parton sub-processes and DGLAP evoluted LO  Partonic Density Functions $f_{i|A}$.
The expression of Eq.~(\ref{minijets})  gives  a mini-jet cross-section  which  rises very fast  with energy.  In order   to ensure unitarity, it is embedded in an eikonal representation,
which requires modeling  the impact parameter space of the colliding hadrons. Neglecting the real part of the eikonal, which at high energy is a good approximation, one then  writes
\begin{equation}
 \sigma _{tot}  = 2\int {d^2 b} \left[ {1 - e^{ - \chi_I(b,s)} } \right]
 \label{sigtot}
\end{equation}
In these models, the imaginary  part of the eikonal function,  $\chi_I$, is calculated from the   average number of collisions as $n(b,s) =2\chi_I(b,s)= n_{soft} (b,s) + n_{hard} (b,s) $, namely  it is split into a soft contribution  which will be parametrized with a suitable non-perturbative expression, and a perturbative (pQCD) term where both hard and soft gluon emission contribute.
 The  mini-jet model  of \cite{ourEPJC}, called Bloch Nordsieck (BN) model,   proposes two different impact parameter  distributions for these two terms,
  i.e.
  \begin{eqnarray}
  n_{soft} (b,s) = A_{FF}(b,s)\sigma _{soft} (s)\label{nsoft} \\
 n_{hard} (b,s) = A_{BN}(b,s)\sigma _{jet} (s)
 \label{nhard}
 \end{eqnarray}
where $A_{FF}$ is obtained from the proton form factor,  and $A_{BN}(b,s)$ as the Fourier transform in impact parameter space of the function $ {d^2P({\bf K_\perp})}/{d^2 {\bf K_\perp}}$ obtained as resummation of all soft gluons, emitted in  parton-parton collisions with $p_t>p_{tmin}$, down into the infrared region \cite{ourproton}, i.e.
\begin{eqnarray}
A_{BN}(b,s)=N \int d^2{\bf  K_{\perp}}\  e^{-i{\bf K_\perp\cdot b}}
 {{d^2P({\bf K_\perp})}\over{d^2 {\bf K_\perp}}}=\\ ={{e^{-h( b,q_{max})}}\over
 {\int d^2{\bf b} \ e^{-h( b,q_{max})}
 }}
\end{eqnarray}
  The function  $h( b,q_{max})$ corresponds to the spectrum for single gluon emission, regularized in the infrared. The extension to infrared gluons is crucial for this model and requires an ansatz as to  the coupling of very soft gluons to the emitting quarks, which the model introduces through a singularity parameter  $1/2<p<1$ \cite{ourproton}.
The  physical content of this model  can be summarized as follows: i) the rise is obtained from low-x parton-parton collision, ii) the taming of the rise is obtained  from
 the Fourier transform in impact parameter space of the
resummation of all soft gluons emitted in the parton-parton collisions.
 The expression for the impact parameter distribution when the two hadrons collide describes  the acollinearity introduced by soft gluon emission which   reduces the  cross-section from mini-jets. The expression  is energy dependent  through  the  parameter $q_{max}$, which embeds the kinematics of single gluon emission, and
represents the maximum transverse momentum for {\it single}  gluon emission and depends on the  $p_t$ of the final state partons 
as detailed in \cite{ourproton} for $pp$ collisions.
Through the ansatz for the coupling in the infrared region, the  distribution $A_{BN}(b,s)$ gives a (logarithmically )  energy dependent cut-off in $\bf b$-space, dynamically generated by soft gluon emission, and reduces the very fast rise from the mini-jet cross section \cite{ourFroissart}.


With the above inputs, and a  parametrization  of the low energy region, i.e. $n_{soft}(b,s)$, the model  gave a good description of
$pp$ and $p{\bar p}$ total cross-sections,   predicting  $\sigma(\sqrt{s}=14\ TeV)= 100 \pm 12 \  mb$, with  the error to
reflect various uncertainties in choice of parton densities and  values for the set of  non-perturbative parameters $\{p,p_{tmin}\}$ \cite{ourproton}.
Following \cite{halzenfletcher}, this  model was then   applied to photoproduction \cite{ourEPJC}, namely
\begin{eqnarray}
 \sigma _{tot}^{\gamma p}  = 2P_{had}\int {d^2 b} \left[ {1 - e^{ - n^{\gamma p}(b,s)/2} } \right]
\label{eq:gamp}\\
P_{had}=\sum_{V=\rho,\omega,\phi}{{4\pi \alpha}\over{f_V^2}}\\
n^{\gamma p}(b,s)=n_{soft}^{\gamma p}(b,s) +n^{\gamma p}_{hard}(b,s)\nonumber \\ = n_{soft}^{\gamma p}(b,s) + A(b,s) \sigma_{jet}^{\gamma p}(s)/P_{had}\\
n_{soft}^{\gamma p}(b,s)= {{2}\over{3}}  n_{soft}^{p p}(b,s)
\end{eqnarray}
The extension to photon process requires the probability $P_{had}$ that the photon behaves like a hadron \cite{collins,halzenfletcher}. This quantity is non perturbative and could have some mild energy dependence. However, to minimize the parameters, it was  taken  to be a constant, estimating it through Vector Meson Dominance.

\begin{figure}
\resizebox{0.5\textwidth}{!}{
\includegraphics{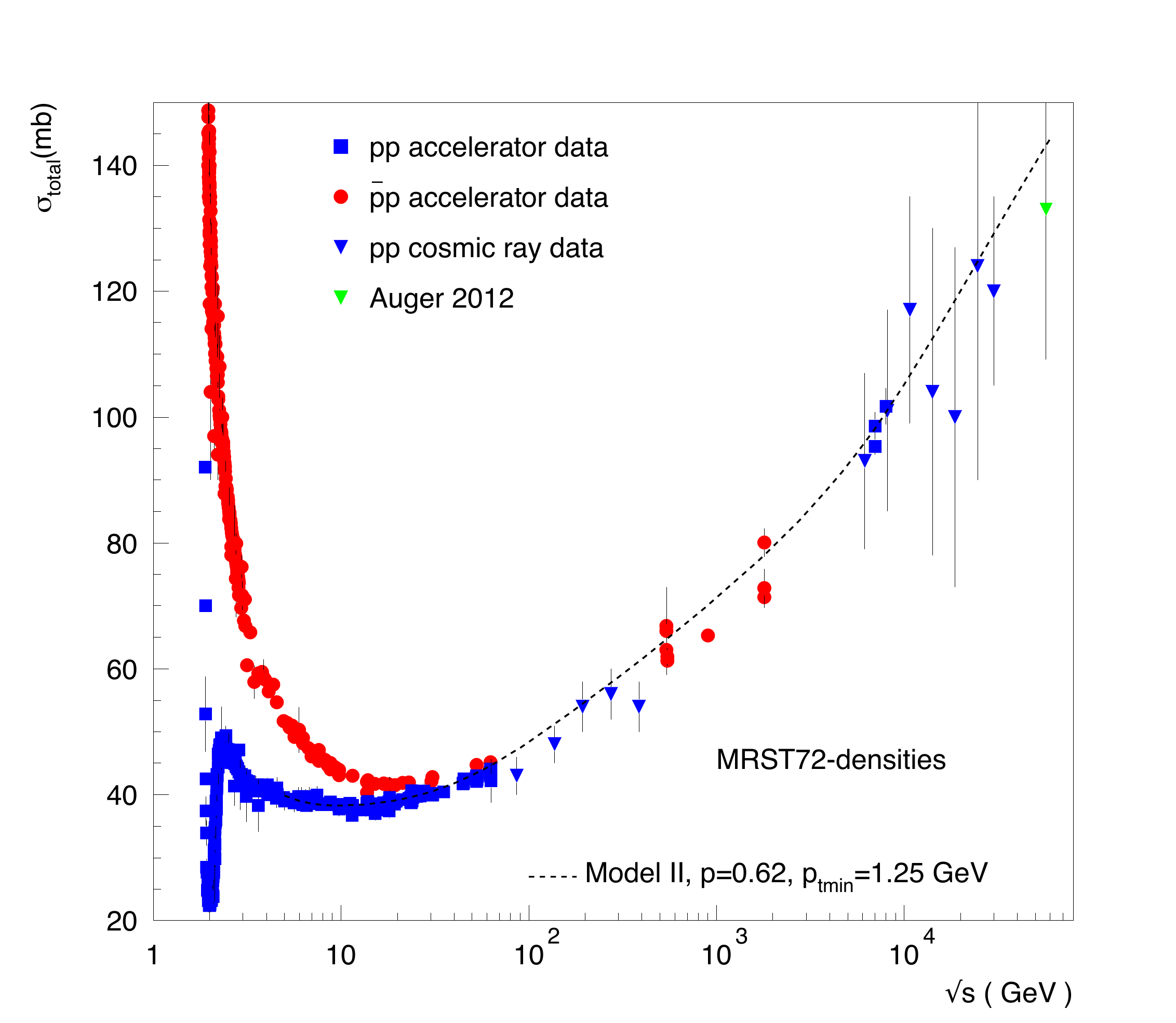}}\\
\caption{Total proton-proton cross-section and its description with the BN model, eikonalized  mini-jets with   soft gluon resummation, described in the text.}
\label{fig:totpp}
\vspace{-1cm}
\resizebox{0.5\textwidth}{!}{
\includegraphics{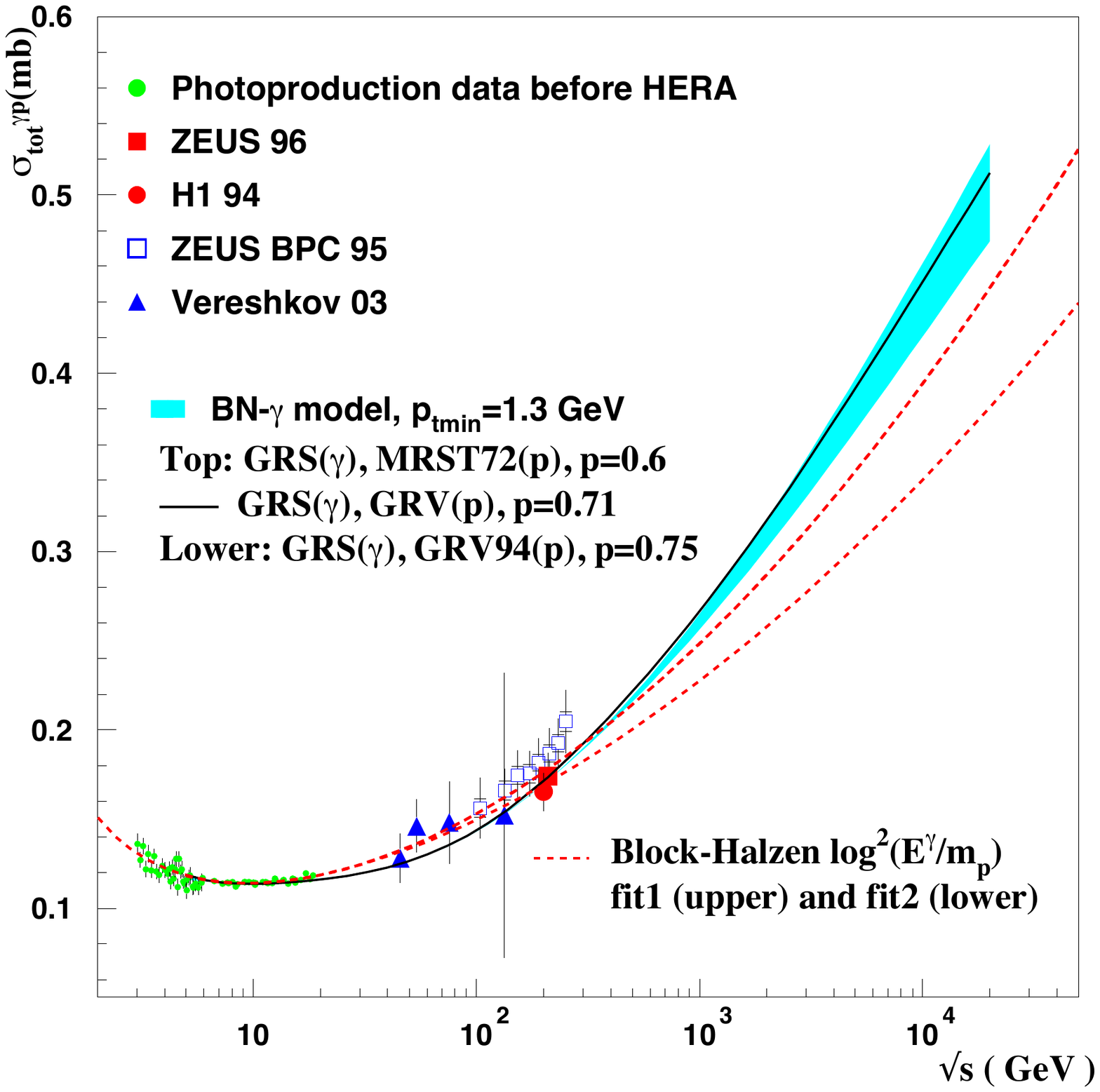}}
\vspace{-2cm}
\caption{Total photoproduction cross-section  and its description with the BN model, and with the  analytical models by Block and collaborators \cite{block2014}.}
\label{fig:gamp}
\end{figure}

With such modifications, one can now calculate the $\gamma p$ cross-section that will be used as  an input in the AIRES  shower simulation program. To determine the parameter sets which  best describe the behavior at energies as high as those reached by cosmic rays,  one should   take into account the impact of recent LHC \cite{TOTEM7,TOTEM8,ATLAStot} and AUGER Observatory \cite{AUGER2012}  results on the $pp$ cross-section, which have appeared after the original analysis of  \cite{ourEPJC}. Indeed, the measurement of the total $pp$ cross-section at LHC, at energies $\sqrt{s}=7$ and $8\ TeV$, has allowed to reduce the systematic uncertainties present in most models, due to the large errors affecting  the ${\bar p}p$ cross-section. We show in Fig. \ref{fig:totpp} how  this model accommodates recent results for
$\sigma^{pp}_{total}$, including the extraction of the $pp$ cross-sections from cosmic ray measurement by the AUGER collaboration, at $\sqrt{s}=57\ TeV$. With the choice of parameters as indicated in the figure and MRST72 densities \cite{MRST},  the value expected at $\sqrt{s}=14\ TeV$ is $\sigma_{total}^{pp}=112.24\ mb$. Using  other  types of LO densities, such as the older GRV \cite{GRV} or more recent MSTW \cite{MSTW},   gives  similar values \cite{ourcosmic},
namely, up to LHC values, the predictions are rather stable, for different densities, once the TOTEM and AUGER points  are  included in the description.

The extension of the model to $\gamma-p$ is shown in Fig.~\ref{fig:gamp}, where GRS are the LO PDFs for the photon \cite{GRS}. In the figure the results from the BN model are compared with results from two fits by Block and collaborators, which impose a Froissart-limit  saturating high energy behavior  \cite{block2014}.
 Before moving to $\gamma-air\ nucleus$ simulations, we notice that, at higher c.m. values where accelerator data are not available, the model we have described gives predictions which differ by 10-20\%, depending on the PDFs used, and also different from both fits of Ref. \cite{block2014}. For our model, the difference corresponds to the different PDFs used, and thus to the different low-x behavior of these densities.


\section{Observables in shower development}
 We now apply the above model to study  atmospheric air showers initiated by photons. For these simulations, we use the  cross-section that corresponds to the solid line within the blue band in Fig. ~\ref{fig:gamp} and fit 2 of ~\cite{block2014}. 
We shall see how  changes in  the $pp$ cross-section at high energy affect the photo-nuclear cross-section at the corresponding energies, and, through this, photon initiated shower observable.  In such showers, the photonuclear reactions constitute the main channel for production
of hadrons. These hadrons are responsible for the production of muons, mainly via pion decay. It is a well known fact
that showers initiated by photons have noticeably less muons than showers initiated by hadrons, and this is one of
the features used to discriminate photon initiated from hadron initiated  showers.

We have performed simulations of extended air showers using the AIRES system \cite{AIRES} together with the hadronic
interaction package QGSJET-II \cite{QGSJET}. We have run two sets of simulations, namely, (1) using the cross sections for
photonuclear reactions at energies greater than 200 GeV that are provided with the currently public version of AIRES;
and (2) replacing those cross sections by the ones corresponding to the present model. We are going to refer to sets
(1) and (2) as {\it old} model and {\it present} model, respectively.

\begin{figure}
\resizebox{0.5\textwidth}{!}{
\includegraphics{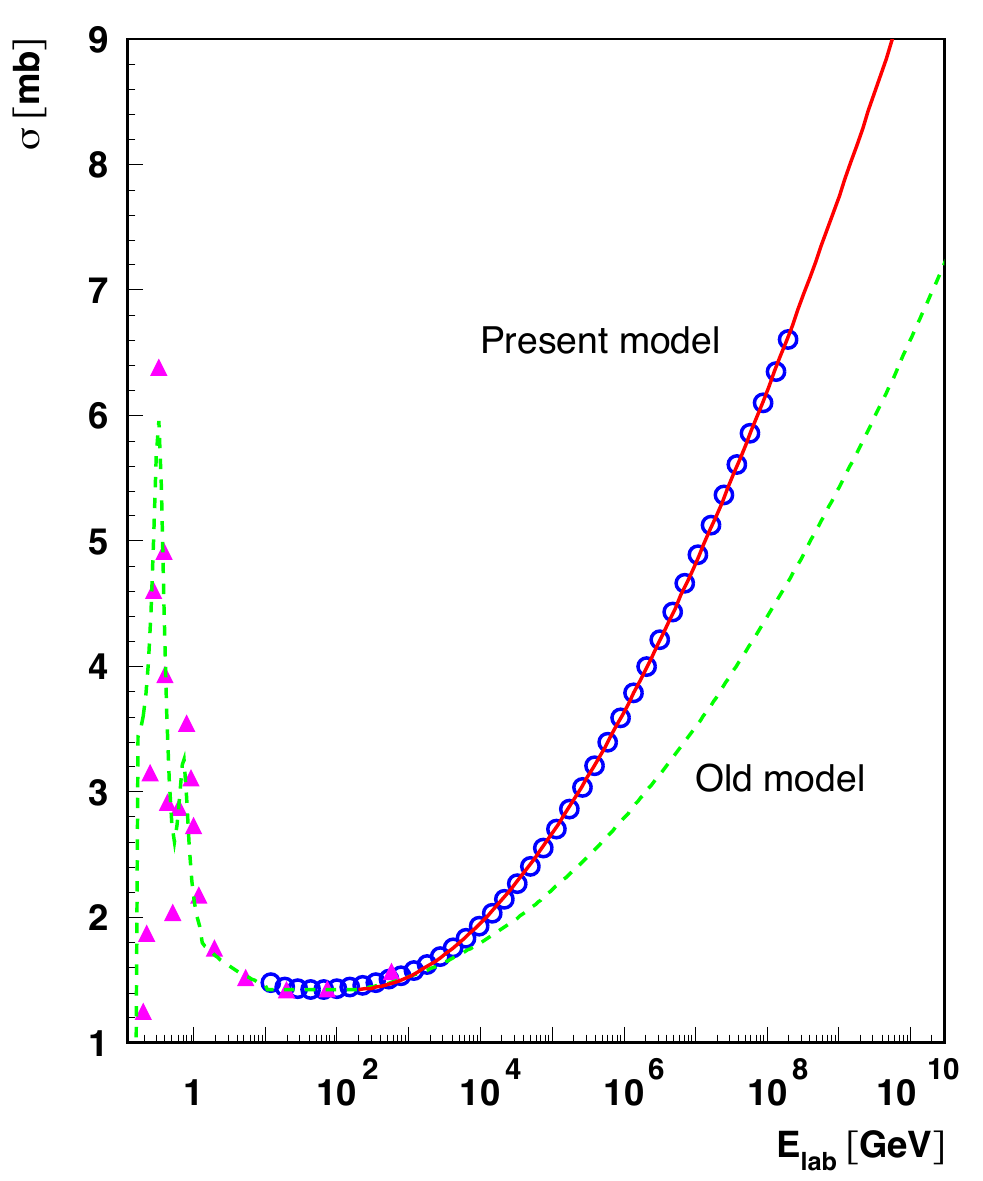}}
\caption{ Photon-air nucleus cross sections used in the present  simulations,
  plotted versus photon lab energy. The triangles correspond to
  experimental data taken from reference \cite{PDG}. The open circles
  correspond to the present model (equation (\ref{eq:gamp})), and the solid
  line corresponds to a fit to these  points, valid for energies greater
  than 200 GeV. The dashed line corresponds to the cross sections used
  in AIRES.}
\label{fig:gammacrsn}
\end{figure}

 In Fig. ~\ref{fig:gammacrsn}  the different $\gamma-air\  nucleus$ cross sections are displayed as a function of photon lab energy.
The triangles correspond to experimental data taken from  \cite{PDG}, while the open circles correspond to numerical
calculations  starting from Eq. (\ref{eq:gamp}). The solid line is a fit to the results from the above  model, valid for energies greater than 200 GeV. The dashed line corresponds to the up to now standard cross sections
implemented in AIRES, the {\it old} model.  Notice that
for energies less than 200 GeV we use always the same cross sections, which are calculated from fits to experimental
data.

\begin{figure}
\resizebox{0.5\textwidth}{!}{
\includegraphics{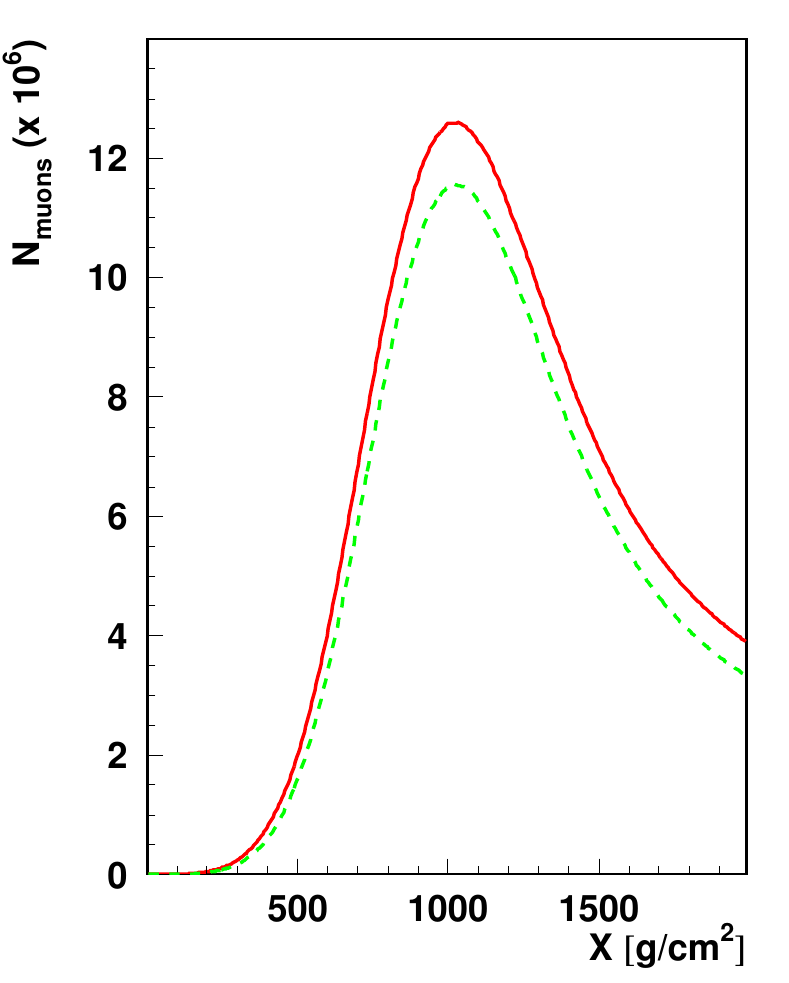}}
\caption{Longitudinal development of muons for
$10^{19}$ eV photon showers inclined 60 degrees. The solid (dashed) line corresponds to
simulations with the present (old) model for photonuclear crosssections.}
\label{fig:longimu}
\end{figure}

\begin{figure}
\resizebox{0.5\textwidth}{!}{
\includegraphics{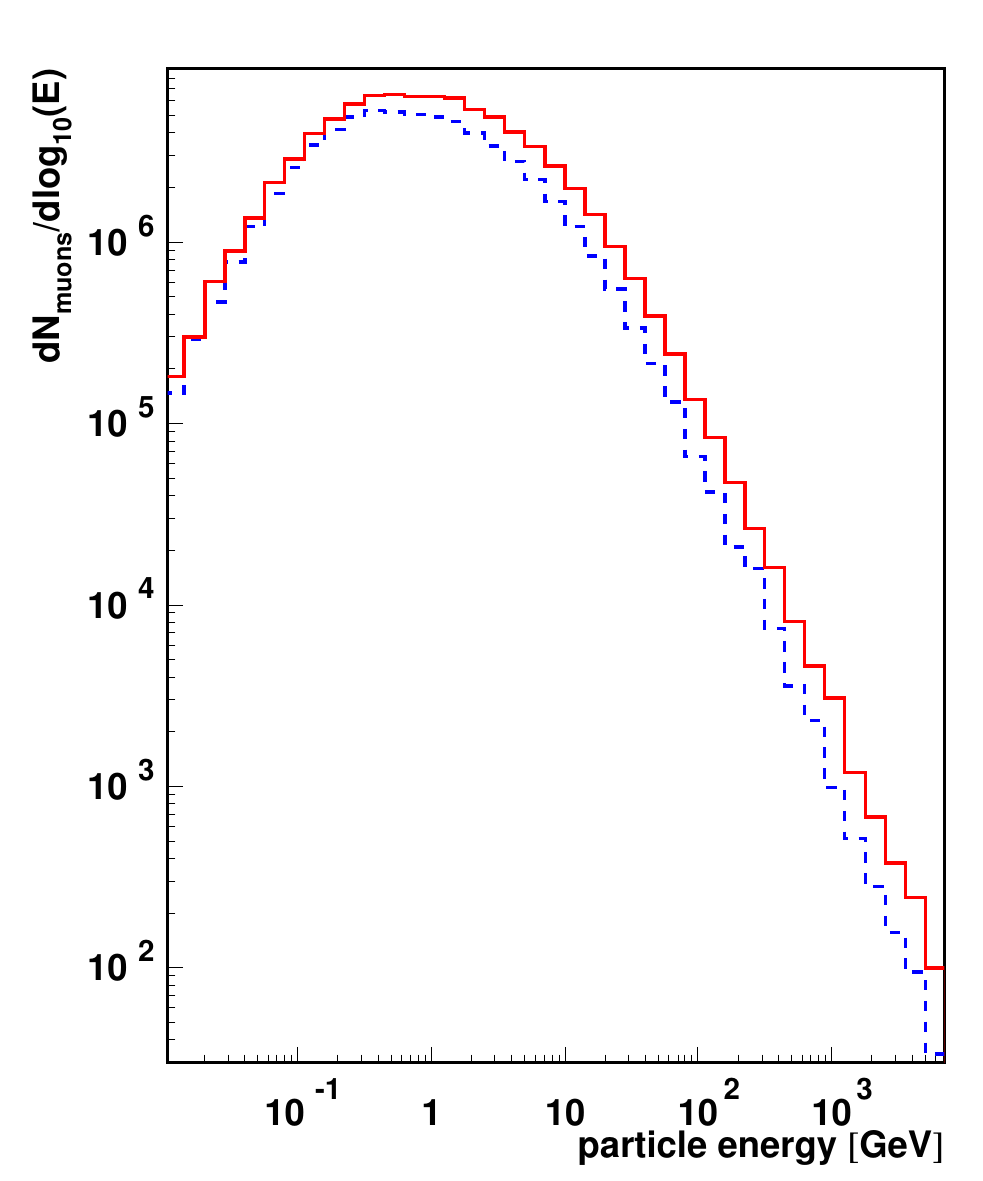}}
\caption{
Energy distribution  of ground muons. The solid (dashed) line corresponds to
  simulations with the present (old) model for photonuclear cross
  sections.}
\label{fig:egydistmu}
\end{figure}

 Fig. \ref{fig:gammacrsn} shows that at energies $E_{LAB}\sim 10^{19}\ eV$ and beyond, there is a large  difference, more than 50\%, between the results for the two models and one can expect this to reflect in some of the observables in the photon initiated showers.  For reasons of brevity, in this contribution  we present results only for this  very representative case, of $10^{19}\  eV$ gamma showers. At this primary energy, geomagnetic conversion is not frequent, thus allowing photons to enter the atmosphere unconverted, and initiate normally the shower development. We have taken in our simulations a ground altitude of
1400 meters above sea level (m.a.s.l.), corresponding to the altitude of current Cosmic Ray Observatories.
The most probable photon interactions at the mentioned energy are electromagnetic (i.e., pair production), and for
that reason most of the shower secondaries will always be electrons and photons; and the number of such secondaries
is not expected to change substantially when replacing the photonuclear cross sections. 
On the other hand, we can expect differences in muon showers, since muons are produced by the decay of unstable hadrons, whose production is affected by the hadronic model for the interaction.  It is also important to consider the characteristics of the muons produced in the simulations. Let us focus on the
representative case of  $10^{19}\  eV$ gamma vertical showers with ground altitude 1400 m.a.s.l. This corresponds roughly
to an atmospheric slant depth of $900\  g/cm^2$.
We present our results for the longitudinal development of muons in Fig.~\ref{fig:longimu}, where it shows up clearly that the simulations with
present model produce more muons in virtually the entire shower life. The relative difference with respect to the old
model is about 12 \% at the maximum ($X \simeq  1100 \ g/cm^2$).
 Accordingly with the results displayed in Fig. ~\ref{fig:longimu} this depth is located
short before the maximum of the muon longitudinal profile.
Concerning the
  lateral distribution of muons, not much difference is noticed between the two models, the shape is very similar, and the results
  differ only in the total number of
particles. On the other hand, the muon energy distributions displayed in Fig.~\ref{fig:egydistmu}
present noticeable differences for muon energies greater than roughly 1 GeV, with the present model giving the largest number of particles at each bin.
For muon energies less than 1 GeV, both distributions are virtually coincident.

Since  shower muons are generated after the decay of unstable hadrons, mainly charged pions and kaons, the enlarged number of muons that show up in Fig. ~\ref{fig:longimu} would be necessarily connected with enlarged hadron production. The results of our simulations agree with this expected behaviour.


\section{Final remarks}
The main objective of this contribution is to present a QCD-based model for photoproduction, updated from the
previous analysis \cite{ourEPJC} in light of recent LHC results for total $pp$ cross-sections, and to study the impact of this
new model on the air shower development. At very high energies, this model produces a photon-air nucleus total cross section significantly
larger than the previous models included in the standard extended air shower studies.

The present analysis based on
simulations by means of the AIRES system clearly shows that for photon initiated showers the total muon production
is increased in a measurable way. This result could be of direct importance in future determinations of bounds for
the highest energy cosmic photon flux. In this respect, a more detailed analysis of this kind of effects is in progress.
\begin{acknowledgement}
This collaboration was partially financed by the Programa de
Cooperaci\'on Cient{\'\i}fico Tecnol\'ogico Argentino-Espa\~nol:
MinCyT-MINCINN.  F.C. and A.G. acknowledge financial support from
Junta de Andaluc{\'\i}a (FQM-330, FQM-101, FQM-6552) and MICINN
projects FPA2010-16696 and Consolider-Ingenio 2010 program CPAN
(CSD2007-00042). Partial support by CONICET and ANPCyT, Argentina is
acknowledged.
\end{acknowledgement}

\end{document}